\documentclass[twocolumn,showpacs,english,aps,pre,superscriptaddress]{revtex4}
\usepackage{graphicx}
\makeatletter
\usepackage{babel}
\makeatother
\begin{document}

\title{Effect of the Centrifugal Force on Domain Chaos in Rayleigh-B\'enard
convection}
\author{Nathan Becker}
\email{nbecker@physics.ucsb.edu}
\affiliation{Department of Physics and iQCD, University of California, Santa Barbara
93106}
\author{J.D. Scheel}
\email{jscheel@its.caltech.edu}
\affiliation{Department of Physics, California Institute of Technology 114-36, Pasadena,
CA 91125}
\author{M.C. Cross}
\email{mcc@caltech.edu}
\affiliation{Department of Physics, California Institute of Technology 114-36, Pasadena,
CA 91125}
\author{Guenter Ahlers}
\email{guenter@physics.ucsb.edu}
\affiliation{Department of Physics and iQCD, University of California, Santa Barbara
93106}

\begin{abstract}
Experiments and simulations from a variety of sample sizes
indicated that the centrifugal force significantly affects rotating Rayleigh-B\'enard convection-patterns.
In a large-aspect-ratio sample, we observed a hybrid
state consisting of domain chaos close to the sample center, surrounded
by an annulus of nearly-stationary nearly-radial rolls populated by occasional defects reminiscent
of undulation chaos. Although the Coriolis force is responsible for
domain chaos, by comparing experiment and simulation we show that the centrifugal force is responsible for the radial
rolls. Furthermore, simulations of the Boussinesq equations for smaller aspect ratios 
neglecting the centrifugal force yielded a domain precession-frequency $f\sim\varepsilon^\mu$ with $\mu\simeq1$
as predicted by the amplitude-equation model for domain chaos, but contradicted by previous experiment. 
Additionally the simulations gave a domain size that was larger
than in the experiment. When the centrifugal force was included in the simulation, $\mu$ and the domain size
closely agreed with experiment.
\end{abstract}

\maketitle
\section{Introduction}

Rayleigh-B\'enard convection (RBC) occurs when the temperature difference
$\Delta T$ across a horizontal layer of fluid
confined between two 
parallel plates and heated from below 
reaches a critical value $\Delta T_{c}$ \cite{SC,BPA00}.
Rotating the sample about a vertical axis at a rate $\Omega>\Omega_{c}$
induces a state of spatio-temporal chaos \cite{KL,GK} known as domain chaos.

\begin{figure}
\begin{center}\includegraphics[%
  width=3.3in,
  keepaspectratio]{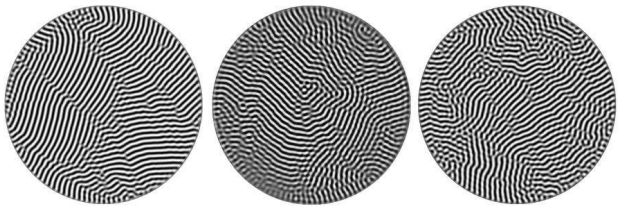}\end{center}
\caption{\label{cap:domain_chaos_images}Images of domain chaos in Rayleigh-B\'enard
convection with $\Gamma=36$, $\varepsilon \equiv\Delta T/\Delta T_{c}-1 = 0.05$, and $\Omega=16.25$.
Left: the temperature profile at the mid-plane from simulation of
the Boussinesq equations with no centrifugal force. Center: shadowgraph
image from experiment. A movie of this sample at $150\times$ actual speed is available \cite{MOV1}. Right: the temperature profile at the mid-plane
from simulation of the Boussinesq equations including centrifugal
force.}
\end{figure}

Domain chaos features patches of straight rolls oriented at various
angles \cite{KL,GK,CB,HB,BH,TC92,CMT94,HEA,HEA2,HPAE}. The orientation of the rolls fluctuates in time and space approximately at discrete
angles due to the Coriolis force which induces the K\"uppers-Lortz
instability.  The size and shape of the domains also fluctuates
and moving defects pepper the entire pattern creating a state of persistent
chaos in both space and time. Figure \ref{cap:domain_chaos_images}
shows several snapshots of domain chaos.

The theoretical models that describe domain chaos usually neglect the           
centrifugal force, relying on the Coriolis                                      
force as the dominant effect brought by rotation.  Even when the full           
Boussinesq equations of motion are considered,                                  
the centrifugal force often is evaluated on the assumption that the density     
is constant throughout the sample \cite{SC}. In that case it has no             
influence on the neutral curve and on the pattern that forms above it.          
In this paper we show, both from experiment and from numerical simulations      
of the Boussinesq equations, that the centrifugal force does have a             
significant influence on the quantitative features of domain chaos. This is     
so even for modest aspect ratios $\Gamma\equiv r_0/d$, where $r_0$ and $d$      
are the radius and height of the convection sample. When $\Gamma$ becomes       
large enough                                                                    
the centrifugal force qualitatively alters the observed patterns. In order      
to reproduce these features in the simulation, it is necessary to consider      
the temperature dependence of the density, as is done explicitly in the         
Appendix.

Qualitatively the influence of the centrifugal force is readily understood.
The vertical density gradient, resulting from the imposed temperature difference, induces a radial large-scale 
circulation (LSC).
The wave director of the RBC rolls tends to align orthogonally to the LSC \cite{CB91}. This tendency competes with the K\"uppers-Lortz mechanism that tends to create disordered and fluctuating domains. 

\begin{figure}
\begin{center}\includegraphics[%
  width=3.3in,
  keepaspectratio]{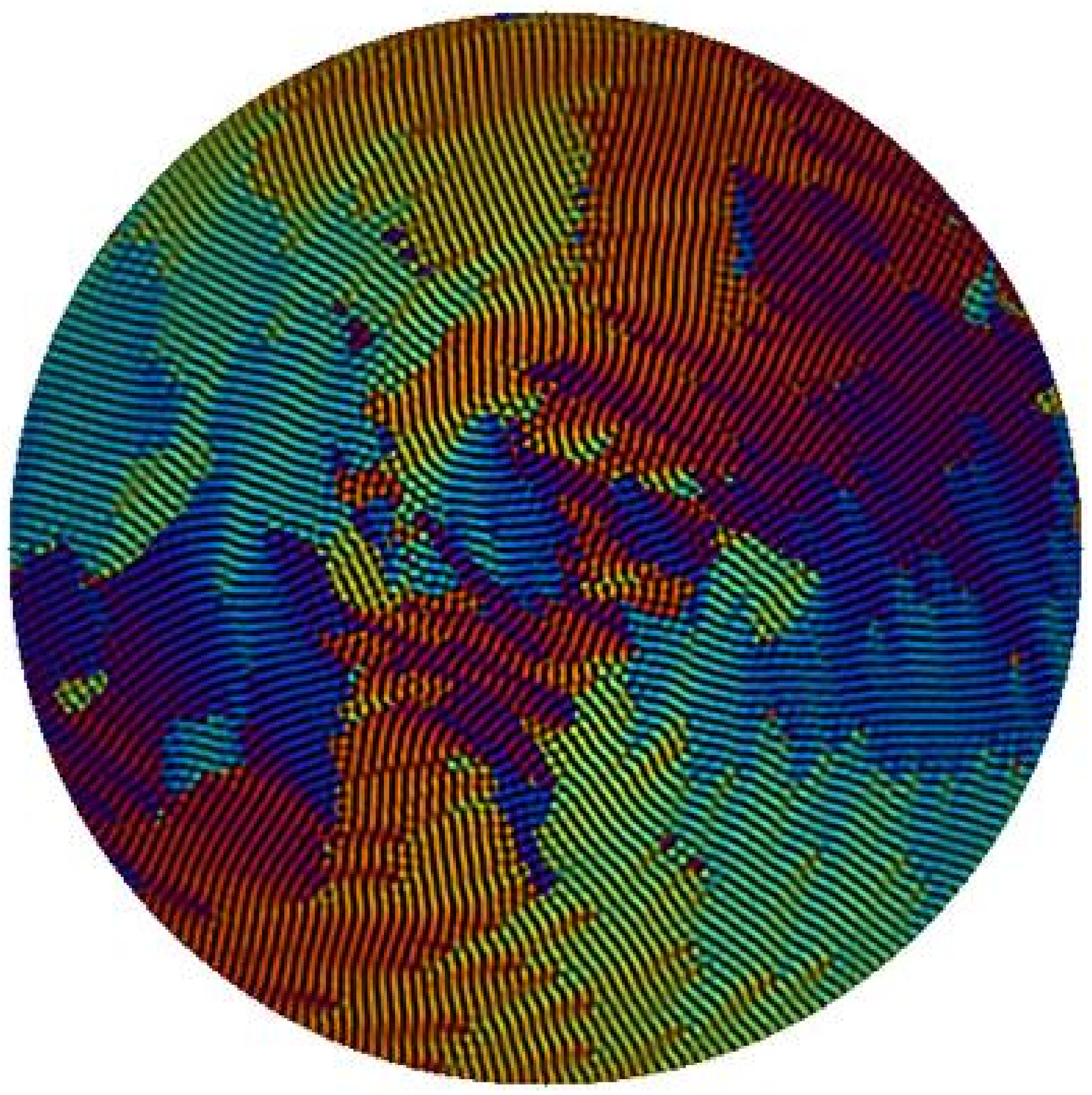}\end{center}

\begin{center}\includegraphics[%
  width=1.5in,
  keepaspectratio]{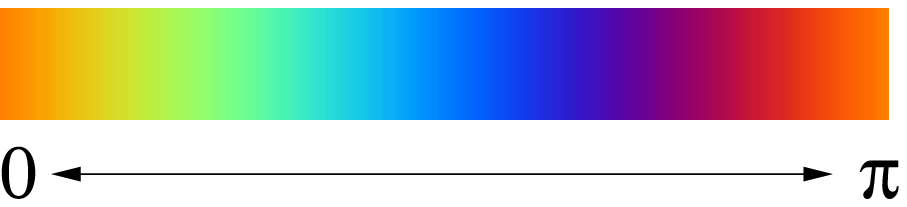}\end{center}

\caption{\label{cap:hybrid_state}(Color online) Hybrid state of domain chaos with radial
rolls and defects along the outer annulus. System parameters are $\Gamma=80$,
$\varepsilon=0.05$, and $\Omega=16.25$. The grey-scale (false color in the electronic version) overlay is
$\theta\left(\mathbf{r}\right)$, the angular component of the local
wave-director field. A movie is available \cite{MOV2} showing the dynamics of this pattern at $150\times$ actual speed.}
\end{figure}

For samples with $\Gamma$ up to about 40 we observed patterns in the experiment that were regarded as typical of domain chaos and similar to those found previously \cite{HEA,HEA2,HPAE}. However, as observed before \cite{HPAE}, the patterns contained domains that were significantly smaller than those of simulations of the Boussinesq equations. At larger $\Gamma=80$ we found experimentally a hybrid state where the
centrifugal force was strong enough relative to the Coriolis force
to qualitatively affect the pattern as shown in Fig.~\ref{cap:hybrid_state}.
There the central section looked like domain chaos, but the annular region
along the perimeter was mostly made up of nearly-stationary nearly-radial rolls.
Defects glided azimuthally across the radial rolls (a movie of this state can be found in the online version). Although simulations of the Boussinesq equations for such large $\Gamma$ were not feasible, the inclusion of an enhanced centrifugal force in simulations for $\Gamma = 20$ showed a similar radial roll structure near the sample boundary.

The outer region of the hybrid state had much in common with undulation chaos \cite{UC,UC2,UC3} observed in
inclined Rayleigh-B\'enard convection. In the inclined system there is
a component of gravity in the plane of the sample.  The
undulation chaos consists of defects gliding across straight
rolls, with the roll axes aligned parallel to this in-plane component. 
Similarly, there is an alignment of the roll axes in
the rotating case that is nearly parallel to the (radial) centrifugal force, with defects gliding azimuthally across the rolls.
The resemblance between these phenomena exists in spite of the fact
that these forces possess a very different character because
the centrifugal force depends on radial position while gravity is uniform throughout the sample.

In the remainder of this paper we first give some experimental and theoretical/numerical details, and then present several results of quantitative pattern analysis that characterize the extent of the influence of the centrifugal force for $15 \leq \Omega \leq 19$ ($\Omega \equiv \omega d^{2}/\nu$ where $\omega$ is the angular rotation rate and $\nu$ the kinematic viscosity) and $\varepsilon \equiv \Delta T / \Delta T_c - 1\alt 0.5$. On the basis of this analysis we show for this parameter range that the crossover from the Coriolis-force-dominated central region of domain chaos to centrifugal-force-dominated near-radial rolls occurs near  $r/d = 35$ ($r$ is the radial coordinate) when $\varepsilon \agt 0.1$, and at somewhat smaller $r/d$ for smaller $\varepsilon$. We conclude that previous samples had been too small to observe the qualitative features of the hybrid state. Unfortunately we also have to conclude that the experimental study of the Coriolis-force effect in compressed gases, where large aspect ratios are attainable, will be severely contaminated by centrifugal-force effects when $\Gamma \agt 40$.

\section{Details of Experiment and Simulation}

The sample exhibiting the large-$\Gamma$ hybrid state (see Fig.~\ref{cap:hybrid_state}) had a height $d = 553~\mu$m with $\Gamma=80$. It contained
sulfur hexa-fluoride  (SF$_6$) at a pressure of 25.44 bars and a mean temperature of
$43.00^{\circ}\textrm{C}$ where the Prandtl number $\sigma\equiv\nu/\kappa$
was 0.88 ($\kappa$ is the thermal diffusivity). We used the shadowgraph technique to image the convection patterns
in an apparatus described elsewhere \cite{EXP}. Although Fig.~\ref{cap:hybrid_state}
shows an image taken at $\Omega = 16.25$, we observed a similar phenomenon
over the entire range $15 < \Omega < 19$ of accessible rotation rates
that were substantially above $\Omega_{c}$.

We also studied a medium and a small sample experimentally that did not exhibit the hybrid state. The medium 
sample had $\Gamma=61.5$, a thickness of 720$\mu$m,
a pressure of 20.00 bars, a mean temperature of $38.00^{\circ}\textrm{C}$,
and $\sigma=0.87$. 
The small sample had $\Gamma=36$, was 1230$\mu$m thick, operating at 12.34 bars, with a mean
temperature of $38.00^{\circ}\textrm{C}$, and $\sigma=0.82$.

In addition to the experiment, we also simulated the Boussinesq equations to study the system theoretically.
A discussion of the relevant equations of motion is given in the Appendix. Details of the numerical code are 
described elsewhere \cite{COMP1,SCHE}. No-slip velocity boundary conditions and conducting lateral thermal
boundary conditions were utilized. We used a spatial resolution of 0.1 and a time resolution of 0.005.
In most simulations that included the centrifugal force, the centrifugal
term was made larger than would be physically realistic, in the present experimental fluid, in order to model 
the effect of a larger aspect ratio.

Two different configurations were used to compute the precession frequency $f$. 
In one case, which included both the centrifugal and Coriolis force, we used
$\Gamma = 20$, $\sigma = 0.93$, $\Omega = 17.6$, and the centrifugal term was twice the physically realistic value
in order to model a $\Gamma$ that would be twice as large.
In the other case, where only the Coriolis force was included \cite{SCHE}, we used
$\Gamma = 40$,  $\sigma = 0.93$, and $\Omega = 17.6$.  
We compare the results of these simulations to the experimental results from Ref.~\cite{HEA}, which also used 
$\Gamma = 40$, $\sigma = 0.93$, and $\Omega = 17.6$.

We also ran some simulations with the exact parameters of the $\Gamma=36$ experimental sample for direct comparison. 
Finally we ran several additional cases for the $\Gamma=20$ simulation using a centrifugal term with
various strengths.

\section{Theoretical considerations}

Although previous theoretical work on domain chaos \cite{TC92,KL,GK,CB,HB,BH,CLM01,CMT94,PPS97,PPS97_2,LPPS}
neglected the centrifugal force because the Froude number $\textrm{Fr}\equiv\omega^{2}r/g$
($g$ is the acceleration of gravity) is small in typical experiments,
other research \cite{HH71,HART} indicated that the Froude number may
not be the only relevant parameter. 
Hart \cite{HART} provided a numerical estimation of the parameter regimes
where the centrifugal force is relevant. He made the approximation
that $\Omega$ is large ($\gtrsim$ 500), so we cannot directly use that estimate because the
present $\Omega$ values are only about 20. We instead consider
\begin{equation}
{\cal A} = \frac {\beta \sigma \Omega \Gamma z_0}{2 u_{0} \varepsilon^{1/2}}
\label{eq:finalratio}
\end{equation} 
($\beta\equiv\alpha\Delta T_c$ where $\alpha$ is the isobaric thermal expansion coefficient). A derivation of the relevant centrifugal and Coriolis terms is given in the Appendix.  
This parameter ${\cal A}$ is the small-$\varepsilon$ approximation to the ratio of the magnitude
of the maximum of the centrifugal term, evaluated near the outer edge of the sample, to the magnitude of the Coriolis term evaluated where it reaches a maximum in the 
horizontal direction,  $z = z_0$. (Note that the temperature of the conduction profile is equal to $-z$ for our system). The quantity ${\cal A}$ 
is an indicator of  the transition from domain chaos to a
hybrid state. We can obtain an approximate numerical value for 
$u_{0}$ by performing a linear stability analysis to get the functional form of the velocity and by using the simulation to get the normalization. For
the case of $\Gamma = 40$, $\sigma = 0.93$ and $\Omega = 17.6$, we
find $u_{0} = 12.3$ at $z_0 = 0.3$. 
These values should also approximately apply for our experimental parameters, which do not vary significantly from the parameters used to obtain the numerical values, except for $\Gamma$ whose
dependence is explicit. For the $\Gamma=80$ sample, $\mathcal{A}=0.61$ for $\varepsilon=0.05$, 
indicating that
the centrifugal force is almost as influential as the Coriolis force, but for
the $\Gamma=36$ sample and $\varepsilon=0.05$, $\mathcal{A}=0.12$ indicating that the Coriolis force
dominates. 
Both samples have similar values of $\Omega$ and
$\sigma$, so it is not surprising that we observe a $\Gamma$-dependent transition to a hybrid
state induced by competition between the centrifugal force and the
Coriolis force.

\section{Results}

\begin{figure}
\begin{center}\includegraphics[%
  width=3.3in,
  keepaspectratio]{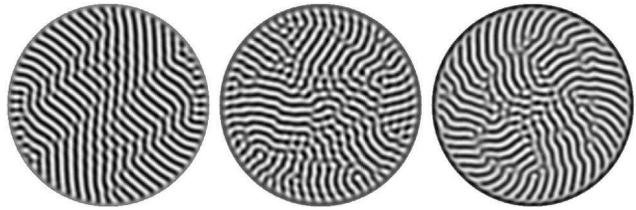}\end{center}

\caption{\label{cap:artificial_sim}Simulation of the Boussinesq equations
with $\Gamma=20$, $\sigma=0.93$, $\Omega=17.6$, $\varepsilon=0.055$,
and an artificially large centrifugal force in order to model the
effect of an inaccessibly large $\Gamma$. Left: zero centrifugal
force. Center: centrifugal force is 4 times the physical value. Right:
centrifugal force is 10 times the physical value.  A movie of this sample at $100\times$ actual speed is available \cite{MOV3}.}
\end{figure}

We qualitatively tested our hypothesis that the centrifugal force
is responsible for the hybrid state by simulating the Boussinesq equations
with centrifugal force included. It was not possible to reach $\Gamma=80$
in the simulation in order to directly reproduce Fig.~\ref{cap:hybrid_state}.
Instead we ran the simulation at $\Gamma=20$ and with the centrifugal
force artificially large in order to model the stronger centrifugal
force at larger $\Gamma$. Some results are shown in Fig.~\ref{cap:artificial_sim}. The qualitative agreement between Fig.~\ref{cap:hybrid_state}
and the right image of  Fig.~\ref{cap:artificial_sim} is striking. Both exhibit domain
chaos at the center but a radial roll structure in an annulus near the side wall. 

In order to make a direct comparison with the experiment we also simulated
the Boussinesq equations for $\Gamma=36$ at $\varepsilon = 0.05$
with the exact parameters of the $\Gamma=36$ experimental sample, both with
and without the centrifugal force included. Figure
\ref{cap:domain_chaos_images} compares examples of the resulting
pattern between simulation, with and without the effect of centrifugal
force, and experiment. As noted before \cite{HPAE}, we observed that the 
domain size was too large when we neglected the centrifugal
force as seen by comparing Figs.~\ref{cap:domain_chaos_images}a and
\ref{cap:domain_chaos_images}b. The domain size in patterns
where we included the centrifugal force, as shown in Fig.~\ref{cap:domain_chaos_images}c,
qualitatively matched the experimentally observed shadowgraph image in Fig.~\ref{cap:domain_chaos_images}b.

\begin{figure}
\begin{center}\includegraphics[%
  width=3.3in,keepaspectratio]{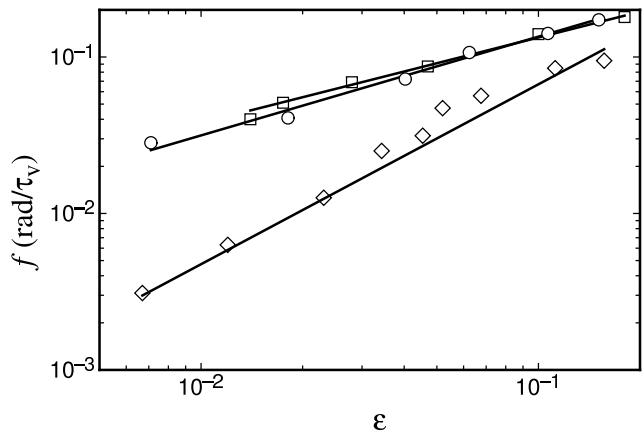}\end{center}
\caption{\label{prescal} 
Effect of the centrifugal force on the scaling of the precession frequency $f$ with $\varepsilon$.  All data are
from samples with $\sigma=0.93$ and $\Omega=17.6$.       	
Circles: from simulations that include both the centrifugal force and the Coriolis force, with $\Gamma=20$ and a
centrifugal term that is twice as large as it would be in the experimental fluid. 
Diamonds: from simulations that include only the Coriolis force, with $\Gamma=40$. Squares: data from previous experiments, with
$\Gamma=40$ \cite{HEA}.
Solid lines: least-squares fits of $f=C\varepsilon^\mu$ to the data.} 
\end{figure}

Recent experiments \cite{HEA} disagreed with the theoretical
prediction \cite{TC92,CMT94} $\xi \sim \varepsilon^{-1/2}$ for the scaling of 
the correlation length $\xi$ and $f\sim \varepsilon^\mu$ with $\mu=1$ for the scaling of the domain precession frequency $f$.
We do not address the $\xi$ scaling in the present work because finite-size effects cloud the issue \cite{CLM01,BA05}.
However, $\mu$ is unaffected by the finite-size effects \cite{CLM01,BA05}.  The method of determining $f$ is
discussed in Refs.~\cite{HEA,SCHE}.
We find that $f$ from the experiment agrees with the simulation when the 
centrifugal force is included. 
The scaling for $f$ is shown in Fig.~\ref{prescal} for simulations both with and without the 
centrifugal force. The simulations that neglected the centrifugal force gave a 
scaling exponent $\mu=1.15$, in good 
agreement with the prediction of $\mu=1$ from amplitude equations \cite{TC92} which also
neglected the centrifugal force. 
There is also good agreement between the experimental data and the results of 
the simulation where both the centrifugal and Coriolis force were included.  
Not only did the numerical simulation yield a scaling exponent of $\mu=0.62$, very close to $\mu=0.58$ from 
the experiment, but it also reproduced the actual values of $f$ in the experiment remarkably well.

Aside from the dramatic effect on the pattern seen in Fig.~\ref{cap:hybrid_state}, in the experiment we also observed a
downward shift by three or four percent of the critical Rayleigh number $R_{c}=\alpha g\Delta T_{c}d^{3}/\kappa\nu$
for the large sample. Previous
work \cite{HH71,HART} showed that the centrifugal force may either stabilize or destabilize
the convective flow depending on the parameter regime, although the
range of parameters specifically corresponding to the present case
was not directly investigated.  However it was suggested that for small $\Omega$ the centrifugal force 
lowers $R_c$ \cite{HART}. 

We did not observe any radial dependence of $R_c$
for the $\Gamma=80$ sample. We measured $R_c$ in both a central circular region of half the sample radius 
and in the outer annulus surrounding
that region.  We accomplished this by computing the time averaged total power in the specified region of the 
experimental images as a function of $\Delta T$.  The total power is proportional to $\mathcal{N}-1$ where
$\mathcal{N}$ is the Nusselt number.  Through a quadratic fit we extrapolated to the background value of the total
power in order to measure $\Delta T_c$, from which we calculated $R_c$ from knowledge of the fluid properties.
The values for $R_c$ in both regions 
agreed within the statistical error of the data, with the biggest difference
being a bit less than 0.2\%, {\it i.e.} an order of magnitude less that the down shift of $R_c$.  

\begin{figure}
\begin{center}\includegraphics[
  width=3in,
  keepaspectratio]{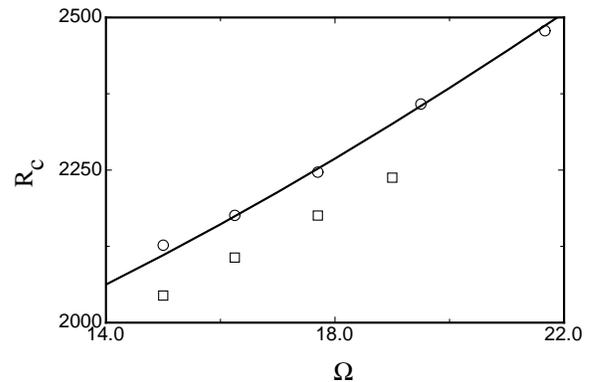}\end{center}

\caption{\label{cap:shiftedRc}Shifted critical Rayleigh number due to the
centrifugal force. Solid line: prediction from linear stability
analysis. Circles: measured from the $\Gamma=61.5$ sample. Squares:
measured from the $\Gamma=80$ sample.}
\end{figure}

The linear stability analysis \cite{SC} for this system was carried out for a
laterally infinite system using the Coriolis force only and neglecting the
centrifugal force. For $\Gamma=61.5$, our measured $R_{c}$ agreed well with
this analysis as is seen in Fig.~\ref{cap:shiftedRc}. Similarly good
agreement had been found earlier for $\Gamma = 40$ \cite{HEA2}. This is
interesting
because we do observe other centrifugal effects for $\Gamma = 61.5$, as
discussed
below. For $\Gamma = 80$ there was a clear downward shift of $R_{c}$,
relative to the theory and to experiments for smaller $\Gamma$, of roughly
3\%.  This trend with $\Gamma$ indicates that the disagreement with theory
cannot be attributed to the finite size of the experimental systems because any finite-size effect on the onset   should have been larger at smaller $\Gamma$. Thus it
suggests that for $\Gamma = 80$ the centrifugal force, which increases with increasing $\Gamma$,  has become strong
enough to affect the onset.

We quantified our analysis of the experimental images by utilizing the angular
component $\theta\left(\mathbf{r}\right)$ of the local wave-director
field computed with an algorithm described in Ref.~\cite{CMT94}, but using higher angular resolution.
Figure \ref{cap:hybrid_state} shows an example of $\theta\left(\mathbf{r}\right)$
 overlayed on top of the shadowgraph pattern in false color.
We used the variance field of $\theta\left(\mathbf{r}\right)$, from
a time series of consecutive images, to address the issue of which
region of the pattern exhibits chaotic dynamics and which region is
mostly dominated by near-stationary radial rolls. Chaotic patterns exhibit
a large variance because the roll orientation fluctuates constantly,
but stationary rolls yield a small variance. 

The mean angular sum field is

\begin{equation}
S\left(\mathbf{r}\right)=\sum_{n}^{N}e^{2i\theta_{n}\left(\mathbf{r}\right)}
\end{equation}

\noindent where $n$ is the image index and $N$ is the number of images. The
complex exponential introduces the necessary periodicity for summing
angles and the factor of 2 treats the field as a director field instead
of as a vector field. The mean angular field is

\begin{equation}
\overline{\theta}\left(\mathbf{r}\right)=\frac{1}{2}\arctan\left(\frac{\textrm{Im}S}{\textrm{Re}S}\right)\textrm{.}
\end{equation}

\noindent The angular variance field is

\begin{equation}
\left\langle \theta^{2}\right\rangle \left(\mathbf{r}\right)=\frac{1}{N}\sum_{n}^{N}\left|e^{2i\theta_{n}\left(\mathbf{r}\right)}-e^{2i\overline{\theta}\left(\mathbf{r}\right)}\right|\textrm{.}
\end{equation}

\begin{figure}
\begin{center}\includegraphics[%
  width=3.3in,
  keepaspectratio]{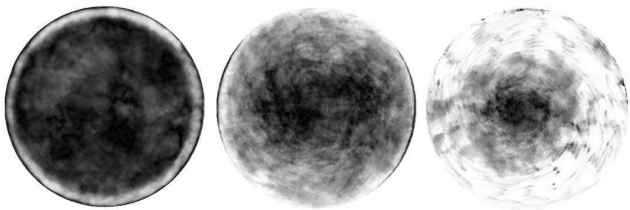}\end{center}

\caption{\label{cap:full_variance_field}The variance field $\left\langle \theta^{2}\right\rangle \left(\mathbf{r}\right)$
for several system sizes from experiment with $\varepsilon=0.05$ and $\Omega=16.25$.
Black corresponds to $\left\langle \theta^{2}\right\rangle \left(\mathbf{r}\right)=2$
and white corresponds to $\left\langle \theta^{2}\right\rangle \left(\mathbf{r}\right)=0$.
Left: $\Gamma=36$. Middle: $\Gamma=61.5$. Right: $\Gamma=80$}
\end{figure}

\noindent Simple algebra shows that $0\leq\left\langle \theta^{2}\right\rangle \left(\mathbf{r}\right)\leq2$
where a value of 0 indicates that the domain orientation is stationary
while a value of 2 indicates that it is maximally fluctuating. The variance
field $\left\langle \theta^{2}\right\rangle \left(\mathbf{r}\right)$
is approximately azimuthally symmetric for several investigated system sizes as
shown in Fig.~\ref{cap:full_variance_field}. There one sees for the larger two samples that the variance is large near the center and smaller near  the side wall.

\begin{figure}
\begin{center}\includegraphics[%
  width=3in,
  keepaspectratio]{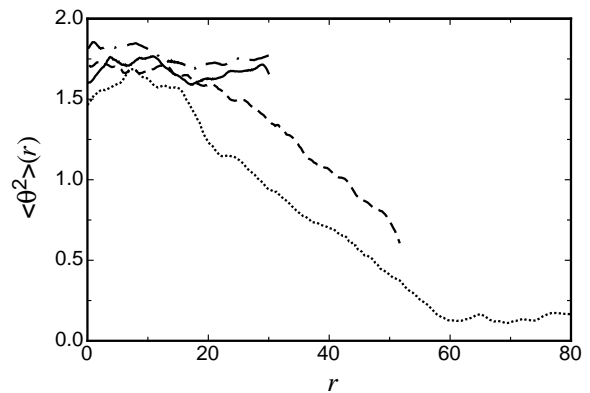}\end{center}

\caption{\label{cap:experiment_theta_variance}Azimuthal average $\left\langle \theta^{2}\right\rangle \left(r\right)$
of the angular variance field (shown in Fig.~\ref{cap:full_variance_field})
for various system sizes from both experiment and simulation. All data were taken with
$\varepsilon=0.05$ and $\Omega=16.25$. Solid line: experiment, $\Gamma=36$.
Dashed line: experiment, $\Gamma=61.5$. Dotted line: experiment, $\Gamma=80$.
Dashed-dotted line: simulation including centrifugal force, $\Gamma=36$.}
\end{figure}

Figure \ref{cap:experiment_theta_variance} displays $\left\langle \theta^{2}\right\rangle \left(r\right)$,
the azimuthal average of the angular variance field, for several system
sizes. For the smallest experimental sample and for a simulation with matching parameters, 
both with $\Gamma=36$, the large variance
across the entire sample revealed the presence of domain chaos throughout.
However, as seen in Fig.~\ref{cap:domain_chaos_images}, even in this case the centrifugal force has reduced the domain size.
Likewise, the central region of the largest sample, with $\Gamma=80$,
exhibited a large variance. Domain chaos dominated near this central
region, but along the perimeter of the sample the variance was very
small indicating coexistence with a more nearly stationary pattern. 
The variance for both the $\Gamma=61.5$ and $\Gamma=80$ samples showed a clear crossover
from domain chaos to radial rolls, however the $\Gamma=61.5$ sample
did not fully transition out of domain chaos even at the edge of
the sample.

\begin{figure}
\begin{center}\includegraphics[%
  width=3in,
  keepaspectratio]{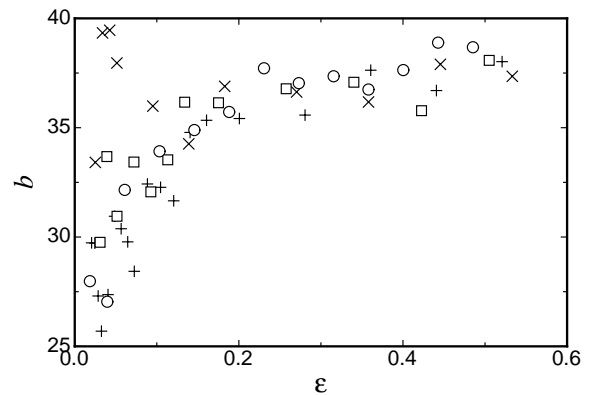}\end{center}

\caption{\label{cap:bwidth}Width $b$ of $\left\langle \theta^{2}\right\rangle \left(r\right)$ for $\Gamma=80$
at various $\varepsilon$. Crosses: $\Omega=15$. Circles: $\Omega=16.25$.
Squares: $\Omega=17.7$. Pluses: $\Omega=19$.}
\end{figure}

We also measured the size of the domain chaos region of the hybrid state by finding the width $b$ of
$\left\langle \theta^{2}\right\rangle \left(r\right)$ chosen such that 
$\left\langle \theta^{2}\right\rangle \left(b\right)\equiv\left\langle \theta^{2}\right\rangle \left(0\right)/2$.
Figure \ref{cap:bwidth} shows the $\varepsilon$ dependence of $b$.  
The behavior of $\mathcal{A}$ (see Eq.~\ref{eq:finalratio}) at small $\varepsilon$ agrees with such 
$\varepsilon$ dependence because of the  factor
 $\varepsilon^{1/2}$ in the denominator, associated
with the Coriolis force, which gives $\mathcal{A}\sim\varepsilon^{-1/2}$ for small $\varepsilon$.
  The data in Fig.~\ref{cap:bwidth} indicate
that, for the present sample parameters, the centrifugal force dominates at small $\varepsilon$ while the 
Coriolis force becomes more important at large $\varepsilon$, although for large enough $\Gamma$ neither 
are negligible.

In addition to looking at the fluctuations in the wave-director angle, we
also measured the angle of the wave director relative to the angle $\theta_{\hat{r}}$ of the side-wall normal
by computing a time-averaged dot-product-like quantity

\begin{equation}
D\left(\mathbf{r}\right)=\frac{1}{N}\sum_{n}^{N}\cos2\left[\theta_{\hat{r}}-\theta\left(
\mathbf{r}\right)\right]\ .
\end{equation}

\noindent Here we are using $\theta_{\hat{r}}$
as a reference for the orientation of the centrifugal force.  The quantity
$D\left(\mathbf{r}\right)$ is 1 when the wave director is parallel to the sidewall normal and
-1 when it is perpendicular.

\begin{figure}
\begin{center}\includegraphics[%
  width=3in]{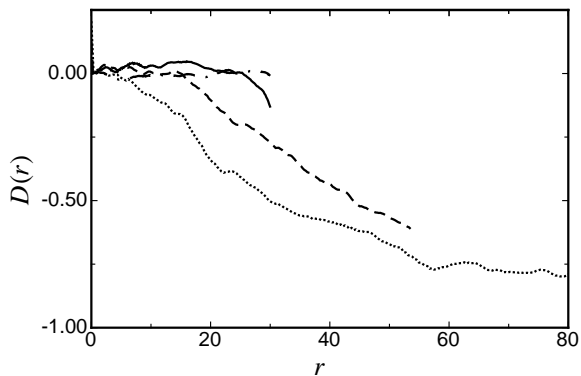}\end{center}

\caption{\label{cap:dotproduct_with_sidewall}The time-averaged dot
product $D\left(r\right)$ (the azimuthal average of  $D\left(\mathbf{r}\right)$)
for the same data as shown in Fig.~\ref{cap:experiment_theta_variance}.
All data were taken with $\varepsilon=0.05$ and $\Omega=16.25$. Solid line: experiment, $\Gamma=36$.
Dashed line: experiment, $\Gamma=61.5$. Dotted line: experiment, $\Gamma=80$.
Dashed-dotted line: simulation including centrifugal force, $\Gamma=36$.}
\end{figure}

In the presence of the LSC induced by the centrifugal force, we expect the wave director
to line up nearly perpendicular to that flow \cite{CB91}.  Some deviation from this might be expected due to the action of the Coriolis force on the Rayleigh-B\'enard rolls. The LSC itself would be expected to nearly align with the centrifugal force and thus should be nearly orthogonal to the side wall. However, here also some deviation from orthogonality to the walls would be expected due to the Coriolis-force influence. As expected, Fig.~\ref{cap:dotproduct_with_sidewall} shows
that, as $\Gamma$ increases, the wave director along the edge of the sample indeed does become more nearly, but not perfectly,  perpendicular
to the sidewall normal.  For the largest $\Gamma$, $D\left(r\right)$ reached a plateau around -0.8, which
corresponded to an angle of about $74^{\circ}$, somewhat shy of a perpendicular angle.

\begin{figure}
\begin{center}\includegraphics[%
  width=3in,
  keepaspectratio]{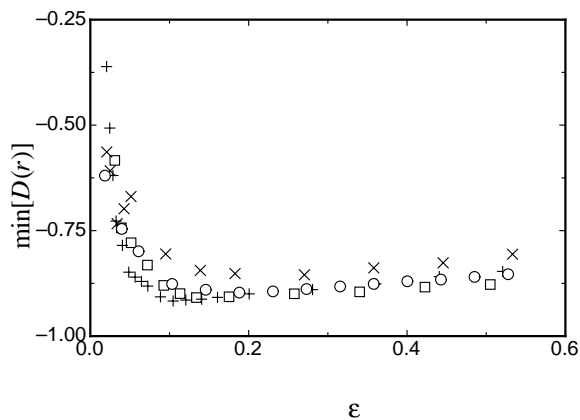}\end{center}

\caption{\label{cap:Dmin}Average minimum $D\left(r\right)$ for $\Gamma=80$
at various $\varepsilon$. Crosses: $\Omega=15$. Circles: $\Omega=16.25$.
Squares: $\Omega=17.7$. Pluses: $\Omega=19$.}
\end{figure}

The angle also depended on $\varepsilon$ as indicated by Fig.~\ref{cap:Dmin},
which shows the minimum value of $D\left(r\right)$ for many curves
similar to those shown in Fig.~\ref{cap:dotproduct_with_sidewall}. To reduce the effect of statistical
fluctuation in the $D\left(r\right)$ curves, the plateau in those
curves was averaged over a small radial length of roughly $5d$ to
yield an average minimum value for $D\left(r\right)$, $\textrm{min}\left[D\left(r\right)\right]$,
at the given $\varepsilon$ and $\Omega$. As a function of $\varepsilon$,
$\textrm{min}\left[D\left(r\right)\right]$ exhibited a minimum, between
$\varepsilon=0.1$ and $\varepsilon=0.3$ depending on $\Omega$,
which corresponded to the nearest realization of orthogonality 
between the sidewall normal and the wave-director. For small $\varepsilon$
the rolls started to turn in and become nearer to concentric rings
along the outer annulus. Such behavior is indicative of the $\varepsilon$
dependence of the relative strengths of the centrifugal and Coriolis
forces.

We also compared auto-correlation functions of the wave-director field between the
experiment and simulation. Using $\theta\left(\mathbf{r}\right)$
we computed

\begin{equation}
C\left(\delta\mathbf{r}\right)=\frac{1}{M\left(\delta\mathbf{r}\right)}\int\cos2\left[\theta\left(\mathbf{r}\right)-\theta\left(\mathbf{r}+\delta\mathbf{r}\right)\right]d\mathbf{r}
\end{equation}

\noindent which is slightly different from the formula used in Ref.~\cite{CMT94}.
The difference is the inclusion of a factor $M\left(\delta\mathbf{r}\right)$,
the number of data points integrated over to find the correlation
at the coordinate $\delta\mathbf{r}$, which corrects for the diminished
quantity of data available at increasingly separated correlations.
This correction results in what is known as an unbiased correlation
function, as opposed to the biased correlation function given when
$M\left(\delta\mathbf{r}\right)=1$. Both approaches have merits,
in particular the biased form must be used for computing power spectra,
but we prefer the unbiased form in the present case.

\begin{figure}
\begin{center}\includegraphics[%
  width=3in]{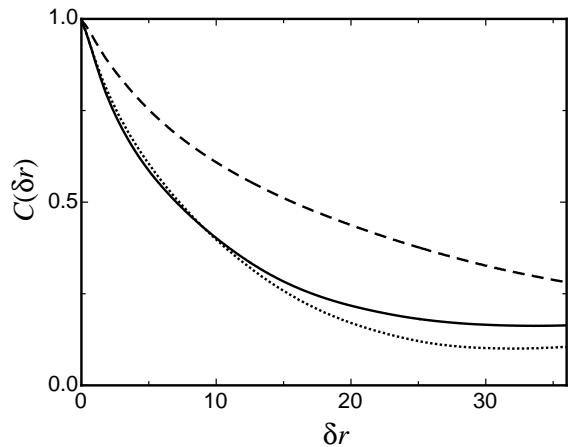}\end{center}
\caption{\label{cap:azim_theta_autocorr}Azimuthal average of the auto-correlation function 
of the wave-director angle field for $\Gamma=36$, $\varepsilon=0.05$,
$\Omega=16.25$, $\sigma=0.82$. Solid line: from simulation of Boussinesq
equations with centrifugal force included. Dashed line: from simulation
but with centrifugal force neglected. Dotted line: from experiment.}
\end{figure}

The angular auto-correlation function magnifies the difference between
centrifugal force and lack of centrifugal force as shown in Fig.~\ref{cap:azim_theta_autocorr}.
The agreement between the experiment and the full simulation with
centrifugal force was excellent at small and moderate $\delta r$, and indicated a quantitative
agreement of the domain size between experiment and simulation with centrifugal force,
as observed qualitatively in Fig.~\ref{cap:domain_chaos_images}.
The slight deviation at large separations should probably be disregarded
due to lack of a sufficient quantity of data, even with the help of
the unbiased form of the auto-correlation function to mitigate this issue.
The simulation which neglected centrifugal force exhibited an overall
larger correlation than the experimental result, in agreement
with the larger domains shown in Fig.~\ref{cap:domain_chaos_images}.

\section{Summary and Conclusion}
We have shown that the centrifugal force plays an important role in the domain chaos state of Rayleigh-B\'enard
convection. Even for relatively small $\Gamma$ where $\mathcal{A}$ remains small throughout the sample, the
centrifugal force affects both the pattern domain size and also the precession frequency.  At large enough $\Gamma$,
a hybrid state forms that contains both domain chaos and radial patterns reminiscent of undulation chaos.  We have
characterized this hybrid state using local wave-director analysis.  We also observed a slight shift in $R_c$ for
this state.

Comparison between simulations and experiment at moderate $\Gamma$ indicated that the centrifugal force is
responsible for the disagreement of previous experiment with the scaling $f\sim\varepsilon$ predicted by the
amplitude model for domain chaos.  A simulation which neglected centrifugal force agreed with that model, but a
simulation which included centrifugal force agreed with the experiment.

\begin{acknowledgments}
We would like to thank Anand Jayaraman and Henry Greenside for some early unpublished work on the centrifugal force effect
on domain chaos as well as useful discussions.  We would also like to thank Werner Pesch for inspiring that early work. This work was supported by the National Science Foundation through Grant DMR02-43336 and by the Engineering Research Program of the Office of Basic Energy Sciences at the Department of Energy, Grants DE-FG03-98ER14891 and DE-FG02-98ER14892.
The numerical code was run on the following supercomputing sites, whom we gratefully acknowledge:
the National Computational Science Alliance under DMR040001 which utilized the NCSA Xeon Linux Supercluster, the  Center for Computational Sciences at Oak Ridge National Laboratory, which is supported by the Office of Science of the Department of Energy under Contract DE-AC05-00OR22725, and
"Jazz," a 350-node computing cluster operated by the Mathematics and Computer Science Division at Argonne National Laboratory as part of its Laboratory Computing Resource Center.

\end{acknowledgments}

\appendix
\section{\label{app:cent} Rotational Corrections to the Boussinesq Equations}
We begin with the Navier-Stokes equation in a rotating frame and in the presence
of gravity, the heat equation and mass conservation:

\begin{eqnarray}
\left(\partial_t + \mathbf u \cdot \nabla\right)\mathbf u  & = & -{\nabla P\over \rho} + \nu\nabla^2\mathbf u - g\hat z - 2 \,\omega\, \hat z \times \mathbf u  + \omega^2 \mathbf r,\nonumber \\
\left(\partial_t + \mathbf u \cdot \nabla\right)T  & = & \kappa\nabla^2 T, \nonumber \\
\nabla \cdot \mathbf u & = & 0,\nonumber \\
\label{navstodim}
\end{eqnarray}
where $\mathbf u$ is the velocity field, $T$ is the  temperature field, $P$ is the pressure field.

We apply the Boussinesq approximation, in which all fluid parameters are assumed to be constant except for the
density in the buoyancy term.  Unlike the standard application of this approximation \cite{SC}, we include buoyancy 
from the centrifugal force as well as gravity.  To lowest order, the density variation in these terms is
\begin{equation}
\rho = \bar \rho[1-\alpha(T - \bar T)],
\label{bouapp}
\end{equation}
where $\bar T$ is the mean temperature and $\bar\rho$ is the density at that temperature.

As long as $\alpha(T - \bar T)$ is small,  we can expand the pressure term:
\begin{equation}
-{\nabla P\over \rho}  \simeq -{\nabla P\over \bar \rho}[1+\alpha(T - \bar T)].
\label{pbou}
\end{equation}
Since the pressure is determined from a gradient, we can absorb the hydrostatic pressure due to the gravitational and the centrifugal forces into the dynamic pressure, by redefining $P  = P+\bar \rho g z \hat z - {\bar\rho\over 2}\omega^2(x^2\hat x + y^2\hat y)$.
The terms proportional to the temperature will not be absorbed. Equation (\ref{navstodim}) becomes:
\begin{eqnarray}
\left(\partial_t + \mathbf u \cdot \nabla\right)\mathbf u  & = & -{\nabla P\over \bar \rho} + \nu\nabla^2\mathbf u + g\alpha(T - \bar T)\hat z \nonumber \\
& - & 2 \,\omega\, \hat z \times \mathbf u  - \omega^2 \alpha(T - \bar T)\mathbf r, 
\label{navstodimbou}
\end{eqnarray}
where we have assumed the term $\nabla  P\alpha(T - \bar T)/\bar \rho$ is 
small.

The variables are then non-dimensionalized by specifying the length in terms of the cell height $d$, the temperature in terms of $\Delta T$, and the time in units of the vertical thermal diffusion time $\tau_{v} = d^2/\kappa$.
 We also define the Prandtl number $\sigma = \nu/\kappa$, and
the Rayleigh number $R= \alpha g \Delta T d^3/\kappa\nu$. In addition, 
we define $\beta = \alpha\Delta T_c$, where  
$\Delta T_c$ (and corresponding critical Rayleigh number $R_c$) is the temperature difference at which conduction gives way to convection.
We obtain:
\begin{eqnarray}
\sigma^{-1}\left(\partial_t + \mathbf u \cdot \nabla\right)\mathbf u & = & -{\nabla P} + \nabla^2\mathbf u\nonumber + R(T - \bar T)\hat z \nonumber \\
& - & 2 \Omega \hat z \times\mathbf u -\beta\sigma\Omega^2 {R\over R_c}(T - \bar T)\mathbf r,\nonumber \\
 \left(\partial_t +  \mathbf u \cdot \nabla\right)T   & = &  \nabla^2 T, \nonumber \\
\nabla \cdot \mathbf u  & = &  0.\nonumber \\
\label{navsto}
\end{eqnarray}
The ratio of the magnitude of the centrifugal term to the magnitude of the Coriolis term is evaluated at the position 
where the Coriolis force reaches its maximum value, which is  where the magnitude of horizontal velocity $u_{\perp}$ is a maximum. To first order
in $\varepsilon^{1/2}$ we can define this magnitude $|u_{\perp}| = u_{0}\varepsilon^{1/2}$, which occurs at $z=z_0$. At this point, the absolute value of $T-\bar T$ = $z_0$, 
the temperature due to the conduction profile. (Note we have neglected convective corrections to the conduction profile, since these are of order $\varepsilon^{1/2}$).
We also want to evaluate the centrifugal force at the horizontal location where it reaches its maximum value, which is at  $r = \Gamma$, 
which then yields Eq.~\ref{eq:finalratio}.

\end{document}